\documentclass[scriptaddress,showkeys]{revtex4-1}
	\usepackage{amsmath}
	\usepackage{makeidx}
	\usepackage{amsfonts}
	\usepackage[ansinew]{inputenc}
	\usepackage[usenames,dvipsnames]{pstricks}
	\usepackage{subfigure}
	\usepackage{epsfig}
	\usepackage{pst-grad} 
	\usepackage{pst-plot} 
	\usepackage[colorlinks,hyperindex]{hyperref}
	\usepackage[squaren]{SIunits}  
\usepackage{multirow}
        
\usepackage{changes}
\usepackage{lipsum}

\begin{document}

\title{Complete Characterization of Stability of Cluster Synchronization in Complex Dynamical Networks}

\author{Francesco Sorrentino$^{1}$, Louis M. Pecora$^{2}$, Aaron M. Hagerstrom$^{3,4}$, Thomas E. Murphy$^{4,5}$, Rajarshi Roy$^{3,4,6}$}
  \affiliation{1. Department of Mechanical Engineering, University of New Mexico, Albuquerque, NM 87131}
  \affiliation{2. U. S. Naval Research Laboratory, Washington, DC 20375}
  \affiliation{3. Department of Physics, University of Maryland, College Park, MD 20742 }
  \affiliation{4. Institute for Research in Electronics and Applied Physics, University of Maryland, College Park, MD 20742}
  \affiliation{5. Department of Electrical and Computer Engineering, University of Maryland, College Park, MD 20742}
  \affiliation{6. Institute for Physical Science and Technology, University of Maryland, College Park, MD 20742}

\date{\today}

\begin{abstract}
Synchronization is an important and prevalent phenomenon in natural and engineered systems. In many dynamical networks, the coupling is balanced or adjusted in order to admit global synchronization, a condition called Laplacian coupling.  Many networks exhibit incomplete synchronization, where two or more clusters of synchronization persist, and computational group theory has recently proved to be valuable in discovering these cluster states based upon the topology of the network.  In the important case of Laplacian coupling, additional synchronization patterns can exist that would not be predicted from the group theory analysis alone.  The understanding of how and when clusters form, merge, and persist is essential for understanding collective dynamics, synchronization, and failure mechanisms of complex networks such as electric power grids, distributed control networks, and autonomous swarming vehicles.  We describe here a method to find and analyze all of the possible cluster synchronization patterns in a Laplacian-coupled network, by applying methods of computational group theory to dynamically-equivalent networks.  We present a general technique to evaluate the stability of each of the dynamically valid cluster synchronization patterns. Our results are validated in an electro-optic experiment on a 5 node network that confirms the synchronization patterns predicted by the theory. 

\end{abstract}

\maketitle

\section{Introduction}

Synchronization of oscillators in large networks has been an interesting problem for many years. It's a phenomenon that shows up in many natural and man-made systems \cite{DeMarco1995, Motter2013, abdollahy2012pnm}. A particular type of synchronization is global synchronization  where the oscillators all follow the same trajectory in state space. A network where this is desirable would be generators in a power grid, as well as some control networks and swarming autonomous vehicles.  Although global synchronization has a well-developed theory \cite{Boccaletti2002, Pecora1998master, PikovskyBOOK2003}  a more recently studied, more complex phenomenon we will call cluster synchronization, which we will abbreviate as CS,  has attracted considerable attention \cite{Zhou2006,Do2012, Dahms2012, Fu2013, Kanter2011, Rosin2013, Sorrentino2007, Williams2013, D'Huys2008, Gabor}.  In CS the network evolves into subsets of oscillators in which members of the same cluster are synchronized to the same trajectory, but members of different clusters are not.  Such synchronized clusters may show up in swarms of animals, where the network is the simple visual link to one's neighbors, or swarms of unmanned autonomous vehicles which are connected by a local communication network. Clusters may also show up in power grids where they would be a sign of a problem, that is, loss of global synchronization.

Given the increase in man-made networks and the growing use of network theory to describe natural systems (e.g., food webs, neuronal and genetic networks) it is important to develop a basic approach to determining what cluster structures are possible in a given network.  In this paper we show what methods can be used to do this using the concept of oscillators coupled through connection to other (not necessarily all) oscillators in the network. More importantly, we extend these methods to the currently unsolved problem of finding clusters in networks of oscillators which have a self-coupling to balance incoming signals from other oscillators, often called Laplacian coupling (more on this below). These allow synchronization clusters that elude other methods to find cluster patterns.  We show how to use and extend symmetry methods to find all possible clusters in such networks of Laplacian-coupled oscillators. We first show how network symmetries can lead to cluster synchronization. We then show how we can go beyond this to analyze CS resulting from Laplacian Coupling.

In Sec. II we review the concepts of symmetries and clusters in networks of coupled oscillators. In Sec. III we discuss methods to uncover all of the possible cluster synchronization patterns in a given network. Our main results are contained in Sec. IV, where we present a stability analysis that applies to any cluster synchronization pattern. 
We also present an elecro-optic experiment that confirms the patterns of synchronization predicted by the theory. Finally, the conclusions are presented in Sec. V.

\begin{figure}[h!]
\centering
\includegraphics[scale=0.7,trim=0.5cm 12cm 0.5cm 0cm,]{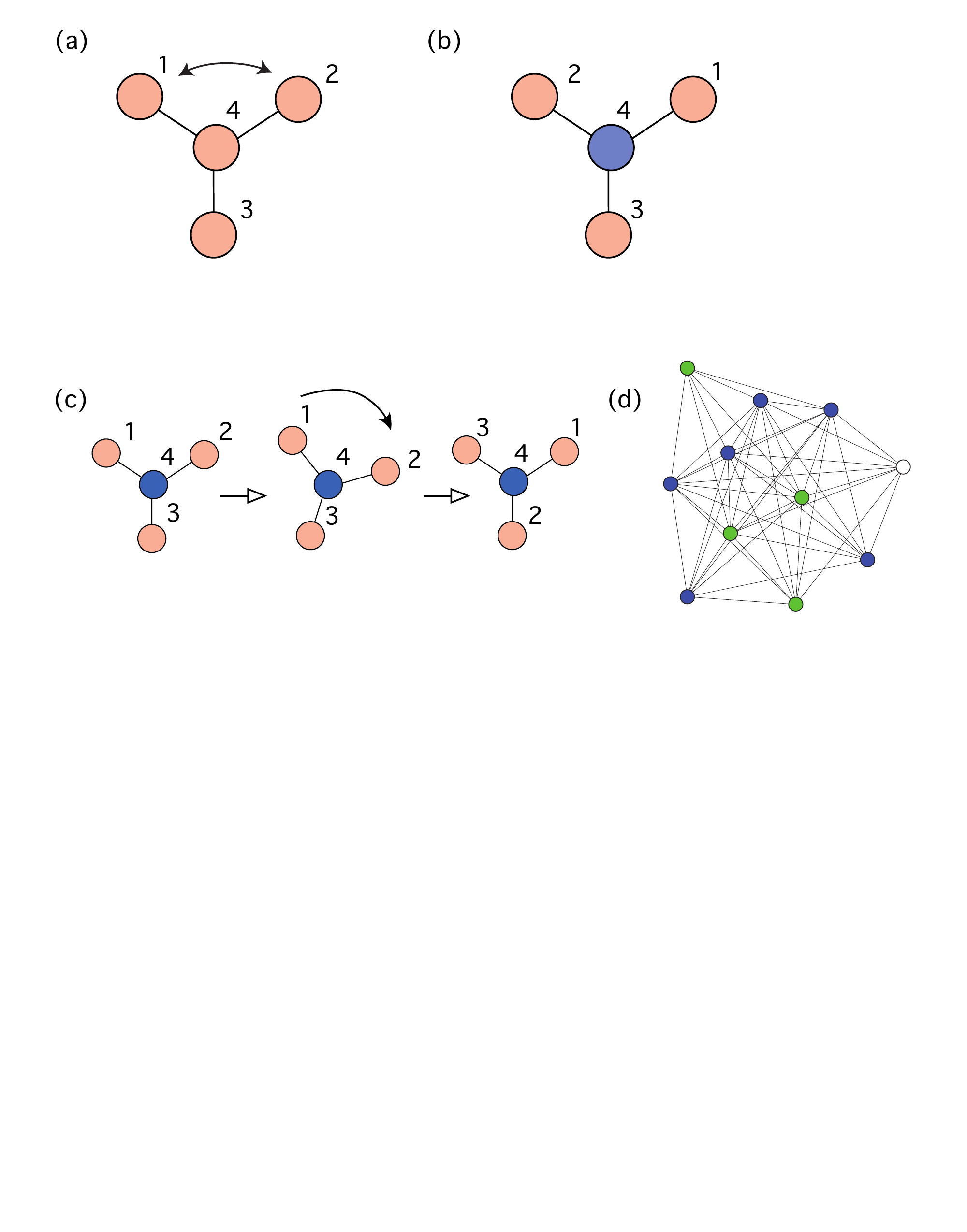}
\caption{(color online) (a) A network of 4 identical oscillators coupled through 3 identical links, (b) the same network after a reflection operation, (c) the same network after a rotation operation, (d) an 11-node network showing 3 clusters (blue, green, white). \label{fig:4node} }
\end{figure}
\section{Symmetries and Clusters in Networks}

Fig.~\ref{fig:4node} (a) shows a 4-oscillator or 4-node network, where the oscillators are identical as are the couplings between them, which are bidirectional, meaning that signals flow in both directions to the oscillators by the same amount and influence on the connected oscillators. This network has a total of 6 symmetries.  We show two.  In Fig~\ref{fig:4node} (b) we show the result of a reflection (shown in (a)) interchanging nodes 1 and 2.  The structure of the network remains the same. In Fig~\ref{fig:4node} (c) we show a rotation of the network by 120{\degree} which also leaves the network indistinguishable from the original in (a). These symmetries are manifest in the symmetries of a matrix that describes the network, the adjacency matrix. This matrix contains a 1 in each position of each row for each node that is connected to the node corresponding to the row number.  For the network in Fig.~\ref{fig:4node} (a) the adjacency matrix $A$ is,

\begin{equation}\label{eq:C}
A=\begin{pmatrix}
  0 & 0 & 0 & 1 \\
  0 & 0 & 0 & 1 \\
  0 & 0 & 0 & 1 \\
  1 & 1 & 1 & 0 \\  
 \end{pmatrix}
\end{equation}

The adjacency matrix plays a crucial role in modeling the dynamics of many networks of identical nodes, as it provides the coupling between the oscillators at each node. We will write the dynamics of the networks as follows,

\begin{equation}
	\label{eq:1}
	\dot{{\bf x}}_i = {\bf F}({\bf x}_i)+\sigma\sum_j A_{ij} {\bf H}({\bf x}_j),
\end{equation}

\noindent
$i=1,...,N$, where ${\bf x}_i$ is the vector of dynamical variables for the $i^{th}$ oscillator, $\dot{{\bf x}}_i$ is the time derivative of the ith node's variables, $N$ is the number of oscillators (nodes),  ${\bf F}$ is the vector field of each node (governing each oscillator's isolated dynamics), and ${\bf H}$ is a coupling function for each link in the network of of each oscillator to another. The entries $A_{ij}$ represent the strength of the coupling from node $j$ to node $i$. Several papers  \cite{belykh2005synchronization, cohen2010dynamic, Kanter2011,  Do2012,  Rosin2013, Belykh2008, D'Huys2008,  Russo2011} have used Eq.\ \eqref{eq:1} to model the dynamics of a network.

 A network symmetry applied to the adjacency matrix leaves it unchanged.  We recall that symmetries of an object (a network in this case) form a mathematical group ${\cal G}$. Any element in this group, say $g$ is represented in the space of network nodes as a permutation matrix, say $R_g$. The invariance of $A$ under the action of the symmetry immediately implies $R_g A =A R_g$, $A$ commutes with all group actions and the equations of motion for nodes that are mapped into each other are the same. For example, in the network in Fig.~\ref{fig:4node} (a), nodes 1,2, and 3 have the same equations of motion, so if they are started from the same initial conditions, they will remain synchronized indefinitely. Node 4 cannot be permuted into any of the other nodes and it will not synchronize with the others, hence we color it differently in the Figure parts (b) and (c). This is the intimate relationship between symmetry and dynamics in networks. We see that it immediately separates the network into two clusters {1,2,3} and {4}.

Many of the recent studies of cluster synchronization are on particular networks that are known or engineered to exhibit cluster synchrony. But as we showed above group theory provides a general approach to finding clusters in arbitrary networks. Steps toward more general approaches using group theory to analyze cluster synchronization began with the work of Golubitsky, Stewart, and Schaeffer \cite{GolubitskySewartBOOK,GolubitskyBOOKII}, where network symmetries are known and can be shown to support CS. Recently, computational methods have been used to study simple symmetric networks and {an approach has been developed that relates the symmetries with the emergence of the CS states} \cite{Nicosia2013}. Finally, we showed that such approaches can be applied to more complex networks with hundreds of oscillators using computational group theory \cite{NC}.

An alternative description of the network dynamics is the following in the case of Laplacian coupling,
\begin{equation}
	\label{eq:1b}
	\dot{{\bf x}}_i = {\bf F}({\bf x}_i)+\sigma\sum_j A_{ij} [{\bf H}({\bf x}_j)-{\bf H}({\bf x}_i)],
\end{equation}
where the coupling from oscillator $j$ to oscillator $i$  is given by the the difference between the output functions ${\bf H}({\bf x}_j)$ and ${\bf H}({\bf x}_i)$. Several papers  \cite{Motter2013, Pecora1998master,  Zhou2006,  Fu2013,  Nicosia2013, Russo2011,ravoori2009adaptive}  have used Eq.\ \eqref{eq:1b} to model the dynamics of a network. Equation \eqref{eq:1b} can be rewritten as follows,
\begin{equation}
	\label{eq:1c}
	\dot{{\bf x}}_i = {\bf F}({\bf x}_i)+\sigma\sum_j L_{ij} {\bf H}({\bf x}_j),
\end{equation}
which has the same structure as Eq. \eqref{eq:1} but now the adjacency matrix has been replaced by the Laplacian matrix, $L=\{L_{ij}\}$,
 $L_{ij}=A_{ij} - \delta_{ij} \sum_j A_{ij}$,  where $\delta_{ij}$ is the Kronecker delta.  By construction then we have that the sums of the rows of the matrix $L$ are equal to zero, i.e. the inputs to the $i$th node are balanced by the diagonal self-coupling.

In Ref.\  \cite{golubitsky2005patterns} Golubitsky, Stewart, and T{\"o}r{\"o}k have shown that for networks that have balanced coupling (all nodes receive the same cumulative input weights, accounting for adjacent nodes and self-coupling, an example of which is Eq. \eqref{eq:1c}), CS can emerge in many patterns that are not directly the result of symmetries. An example of this is global synchronization, which is not  a result of symmetries in the network.  

In general, the patterns of cluster synchrony that can be observed in a network are not unique \cite{belykh2000hierarchy,belykh2001cluster}, hence an important problem is that of determining the parameter ranges for stability and multi-stability for the observed patterns. While it is known that stability of the global synchronization  state for an arbitrary network can be characterized by using the master stability function formalism \cite{Pe:Ca}, a corresponding analysis that applies to the CS patterns that may emerge in a network is not available. In this paper we address this problem by providing necessary and sufficient conditions for stability of each individual CS pattern under very general assumptions. Our analysis applies to systems for which the functions describing the individual dynamics  (possibly chaotic) and the interactions between the systems are arbitrary and to both the cases that the network is described by an adjacency matrix (Eq.\ \eqref{eq:1}) or by a Laplacian matrix (Eq.\ \eqref{eq:1c}), both of which are used to model network interactions.


In what follows we show how to find {\em{all}} of the CS patterns that may emerge in a given network topology, described either by an adjacency matrix or by a Laplacian matrix  and how to evaluate the stability of each allowed pattern.  
We demonstrate all the above phenomena in an electro-optic experiment on a 5 node network that displays all of the possible CS patterns predicted by the theory. 

\section{Analyzing Cluster Synchronization Patterns}

Here we attempt at addressing the following problem: given a network structure (either in terms of an adjacency matrix in Eq.\ \eqref{eq:1} or of a Laplacian matrix in Eq.\ \eqref{eq:1c}) can we find all of the cluster synchronization (CS) patterns that are allowed? For simplicity, we will proceed under the assumption that the network dynamics is described by Eq.\ \eqref{eq:1c} but all of our results include the simpler case that the dynamics is described by Eq.\ \eqref{eq:1}. Indeed, as we will see, the case of the Laplacian matrix is in general more complex to deal with than that of the adjacency matrix. So we consider the most difficult case.

First of all, we note that a symmetry of the adjacency matrix is also a symmetry of the corresponding Laplacian matrix and viceversa \cite{NC}.
Suppose ${\cal G}$ is a group of permutations of the nodes of the network which leaves the coupling matrix $L$ invariant. Then for each $g\in {\cal G}$ we have a permutation matrix $R_g$ that operates on the set of all node vectors ${\bf x}=({\bf x}_1...,{\bf x}_N)^T$. Since $R_g L= L R_g$, this means $d{(R_g{\bf x}}_i)/dt = {\bf F}(R_g{\bf x}_i)+\sigma\sum_j L_{ij} {\bf H}(R_g{\bf x}_j)$, i.e., the symmetry operation leaves the equations of motion unchanged. Hence, the subset of nodes permuted among each other by the group will remain synchronized if started in a synchronized state. We will refer to these subsets of nodes as clusters. The synchronized states for each cluster are flow invariant. 

 An approach to the construction of all allowed  CS patterns in a network has been proposed in \cite{KameiCock}. While the method presented in \cite{KameiCock} is general, as it applies to any network topology, it is computationally expensive. Here we argue that for the case of symmetric networks, a faster approach can  often be followed that takes advantage of computational group theory, which is quite efficient.  At each step we show the results of this method applied to a particular case.

We first decompose the symmetry group ${\cal G}$ into subgroups ${\cal H}_i, i=1,...,\nu$, each of which acts only on some subset of clusters (often only one), but not on any of the others \cite{MacArthur2009,MacArthur2008}. For this reason, we will refer to these subgroups as \emph{cluster groups} and the original group is a direct product of all cluster groups ${\cal G}={\cal H}_1 \times ...  \times{\cal H}_\nu $.  We further decompose each cluster group ${\cal H}_i$ into all of its possible subgroups, which will give us a natural set of symmetry-breaking paths.  These subgroups provide the full range of possible symmetry clusters from the original full symmetry clusters to subclusters to the trivial case where each node is in its own cluster, i.e. no symmetries.      While for the case of the adjacency matrix, this allows one to find all of the possible CS patterns, in the case of the Laplacian matrix, these patterns are certainly valid but there may be other valid patterns that are not predicted by the computational group theory analysis. Extra steps are thus required in order to find these additional patterns.

\begin{figure}[!ht]
\centering
\includegraphics[scale=1]{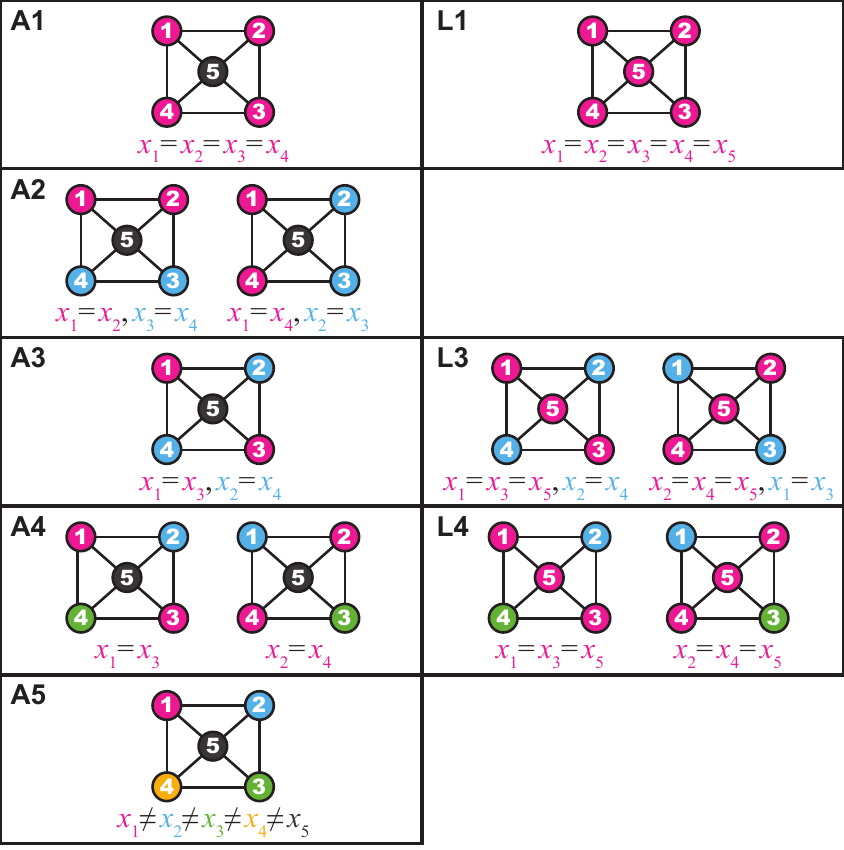}
\caption{(color online) On the left: \textbf{all possible} patterns displayed when the network connectivity is given by the adjacency matrix \eqref{eq:A}. On the right: additional patterns displayed when the network connectivity is given by the Laplacian matrix \eqref{eq:L}. \label{fig:5node} }
\end{figure}

We examine a particular case, a $5$-node network (Fig.~\ref{fig:5node}), with adjacency matrix 
\begin{equation}\label{eq:A}
A=\begin{pmatrix}
  0 & 1  & 0 & 1 & 1 \\
  1 &0  &1 & 0 & 1  \\
   0 &1  & 0& 1 & 1  \\
 1 &0  & 1 & 0 & 1  \\  
   1 & 1  & 1 & 1 & 0  
 \end{pmatrix}
\end{equation} 
and Laplacian matrix
\begin{equation}\label{eq:L}
L=\begin{pmatrix}
  -3 & 1  & 0 & 1 & 1 \\
  1 &-3  &1 & 0 & 1  \\
   0 &1  & -3 & 1 & 1  \\
 1 &0  & 1 & -3 & 1  \\  
   1 & 1  & 1 & 1 & -4  
 \end{pmatrix}
\end{equation} 

\noindent
Figure \ref{fig:5node} shows all the allowed patterns, where nodes belonging to the same cluster are colored the same. 

Cluster patterns A1 to A5 in Fig.~\ref{fig:5node} can be found by performing an analysis of the group ${\cal{G}}$ and all of its subgroups. These are {all the patterns that may emerge for the adjacency matrix \eqref{eq:A} and} the only patterns that may emerge by symmetry for the Laplacian matrix \eqref{eq:L}. The orbits of the original symmetry group  are $\{1, 2, 3, 4 \}$ and  $\{5\}$ by itself and are associated with two cluster groups: ${\cal H}_1$ which permutes the first  and ${\cal H}_2$ which is only the identity for the 2nd single node cluster. Other patterns (A2-A5) are possible based on subgroups of the original group.  These are the result of symmetry breaking bifurcations. 

Now we create new potential clusters by first, choosing a set of cluster groups (${\cal H}_i$) and/or their subgroups (one for each cluster group).  Together these determine a subgroup ${\cal G'}$ of the original group ${\cal G}$.  Second, we combine or merge some of these  clusters as candidates for new synchronized clusters (it's our choice which to try). Because we are merging clusters or subclusters from different original symmetry clusters,  the resulting CS patterns will not be the result of symmetries but may be dynamically valid when the coupling is Laplacian.  Third, we can set those dynamical variables ${\bf x}_i$ equal to others in the merged cluster and see if their equations of motion are the same (guaranteeing flow invariance). However, examining equations of motion by eye for large networks can become prohibitive.  

There is a more direct way of doing step three above using the power and efficiency of group theory and computational discrete algebra tools \cite{GAP4,Stein}. Examining the coupling term in Eq.~(\ref{eq:1c}) when clusters are synchronized the diagonal feedback term (${\bf H}({\bf x}_i)$) of a node will cancel the coupling terms of nodes from the same merged cluster. Hence, in the synchronized state the dynamics behaves as though nodes from the same merged cluster are not connected. We therefore form a {\em dynamically equivalent} coupling matrix which is the original Laplacian matrix with off-diagonal components between nodes from merged clusters set to 0 and the diagonal values then set to the negative of the new row sums. We then perform the cluster group decompositions and subgroup constructions on the new, dynamically equivalent Laplacian (note that if the original Laplacian matrix is symmetric, so are also the dynamically equivalent matrices obtained through this construction). If some set of subgroups of this new coupling matrix has symmetries yielding clusters which are our merged clusters, then their dynamics is flow invariant in the synchronized state and the cluster merging is possible. In this case we call the new clusters {\em Laplacian clusters}.  All the above can be automated in software. 
Hence, in Laplacian networks, even when a CS pattern is not the direct result of a symmetry of the original coupling matrix, it is the result of a symmetry of a dynamically equivalent coupling matrix. This is particularly relevant in terms of computational complexity. In fact, while the problem of finding all of the symmetry operations of a given matrix has not been proven to be polynomial, efficient discrete algebra routines have been developed that make these computations possible even for very large networks (e.g., the Internet at the AS level, for which $N=22332$, see Ref. \cite{MacArthur2008} ). This is in contrast with the method proposed in  \cite{KameiCock}, not based on evaluation of the symmetries, which has been shown to become inefficient even for networks of moderate size, e.g., $N=15$. 

In Fig.~\ref{fig:5node} the only possibility of merging is for node $5$ to be absorbed into one of the remaining clusters in patterns $A_1,A_2,...,A_5$, which is shown on the right hand side. Analyzing our dynamically equivalent matrix as above, we find three additional allowed CS patterns, $L_1$, $L_3$, and $L_4$. These are not predicted by the original symmetry analysis but can emerge when the coupling matrix is Laplacian. Note that the potential $L_2$ patterns from merging of the center with one of the other clusters in A2 will {\em not} lead to a new synchronized cluster (the center has two inputs from the unsynchronized cluster while the corners have only one). The group analysis of the dynamically equivalent network agrees with this.

To conclude, when the network is described by an adjacency matrix, a full characterization of all the dynamically valid patterns can be obtained by taking advantage of available computational discrete algebra tools \cite{GAP4,Stein}. These output the cluster groups,  ${\cal H}_1$, ${\cal H}_2$,... ${\cal H}_\nu$ and a decomposition of each cluster group ${\cal H}_i$ into all of its possible subgroups, which provides a natural set of symmetry-breaking paths.
As discussed above, available computational group theory routines perform these tasks very efficiently, when compared to other possible methods (e.g., \cite{KameiCock}). For this case our approach is always better than the state of the art.
When the coupling is in Laplacian form, all of the (symmetry-related) CS patterns of the associated adjacency matrix are mantained. Moreover, it is possible for some of the cluster groups  ${\cal H}_1$, ${\cal H}_2$,... ${\cal H}_\nu$ and some of their subgroups to merge to form new dynamically valid clusters. In order to find all of the dynamically valid mergings, it can be helpful to use the clusters and subclusters provided by the symmetry analysis as the building blocks of our algorithm. If the number of these different clusters and subclusters is equal to $\mu$, the number of tests that need to be performed (for pairs, triplets, and so on) is upper bounded by  $B_\mu$, the $\mu$-th Bell number, hence it grows combinatorially with $\mu$. For all the networks that do not display many symmetries, i.e., for which $\mu \ll N$, our approach based on computational group theory will be much faster than the one presented in \cite{KameiCock}. Hence, in general it is a good idea to preliminarily run an analysis of the symmetries of the network to assess whether  $\mu < N$ and choose the most convenient approach based on this outcome. It should be noted that there is no need for testing for mergings between subgroups of the same group, which reduces the complexity of the calculations. In general, finding all of the dynamically valid patterns for Laplacian networks may be substantially harder than for networks described by a symmetric adjacency matrix. Similar limitations were observed in \cite{KameiCock}.

\section{Stability Analysis and experimental validation}

Stability of cluster synchronization has been investigated for phase oscillators \cite{Nicosia2013} and Stuart-Landau oscillators \cite{poel2015partial}, and for lattices of coupled systems  \cite{belykh2000hierarchy,belykh2001cluster}. However,
a general approach to analyze stability and multistability of CS patterns in arbitrary networks has not been developed.  
In \cite{NC} we studied a particular CS pattern corresponding to a minimum number of clusters (i.e., maximal symmetry) for the case that the connectivity of the network is in the form of an adjacency matrix.

We now develop variational equations for the merged-cluster system so we can calculate the stability of  each one of the allowed CS patterns. While a number of papers \cite{manrubia1999mutual,belykh2000hierarchy,belykh2001cluster,belykh2011mesoscale,golubitsky2005patterns,Nicosia2013,poel2015partial} have dealt with cluster synchronization in networks, only Refs.\  \cite{Nicosia2013,poel2015partial} have considered the problem of stability for particular dynamics of the individual systems. Refs. \cite{belykh2000hierarchy,belykh2001cluster,belykh2011mesoscale} have emphasized that for arbitrary systems this is a difficult problem. In order to analyze stability, we start with the subgroup ${\cal G'}$ of the original group which generated the clusters we want to merge.  It is formed by a direct product of the subgroups we choose to use in our merged system. Using ${\mathcal C}_m$ to represent each cluster of nodes, $m=1,...,K$, where $K$=number of clusters in ${\cal G'}$, we have the variational equation of Eq. (\ref{eq:1}),

\begin{eqnarray}\label{linv}\nonumber
\delta \dot{{\bf x}}(t) &=& \left[\sum_{m=1}^{M} E^{(m)} \otimes D{\bf F}({\bf s}_m(t)) \right.  \\ 
&&  \left. + \sigma \sum_{m=1}^{M} ( LE^{(m)}) \otimes D{\bf H}({\bf s}_m(t)) \right] \delta {{\bf x}}(t),
\end{eqnarray}
where the $Nn$-dimensional vector $\delta {\bf x}(t)=[\delta {\bf x}_1(t)^T,\delta {\bf x}_2(t)^T,...,\delta {\bf x}_N(t)^T]^T$, $L$= the Laplacian-coupling matrix, and $E^{(m)}$ is an $N$-dimensional diagonal \emph{indicator matrix for each cluster} such that
$E^{(m)}_{ii}$ is  equal to $1$ if  ${i \in {\mathcal C}_m}$ and is equal to  ${0,}$  otherwise, $i=1,...,N$.  Note that we must use the original Laplacian matrix and not the dynamically equivalent one, which is used only for detecting synchronization flow invariance.

\begin{table}
\begin{center}
    \begin{tabular}{ | l | l |  l |}
    \hline
    Pattern & Quotient Network Dynamics (right hand side is $\text{mod} 2 \pi$) & Transverse Perturbations \\ \hline
\multirow{2}{*} {A1:} & $x^{t+1}_{1}=\left[ (\beta-\sigma) {\cal I}(x^{t}_{1}) + \sigma {\cal I}(x^{t}_{5}) + \delta \right]$   & $\eta^{t+1}= \left[ \beta  + \sigma \lambda  \right] {D\cal I}(x^{t}_{1}) \eta^t$, $\lambda=-3,-5$ \\
{$x_1=x_2=x_3=x_4$} & $x^{t+1}_{5}=\left[ (\beta- 4 \sigma) {\cal I}(x^{t}_{5}) + 4 \sigma {\cal I}(x^{t}_{1}) + \delta \right]$ & \\
 \hline
   \multirow{3}{*} {A2:}
&  $x^{t+1}_{1}=\left[ (\beta- 2 \sigma) {\cal I}(x^{t}_{1}) + \sigma {\cal I}(x^{t}_{4})+ \sigma {\cal I}(x^{t}_{5}) + \delta \right]$   & $\eta_1^{t+1}= \left[ \beta  -4 \sigma  \right] {D\cal I}(x^{t}_{1}) \eta_1^t - \sigma {D\cal I}(x^{t}_{3}) \eta_3^t$  \\
  & $x^{t+1}_{2}=\left[ (\beta- 3 \sigma) {\cal I}(x^{t}_{2}) + 3 \sigma {\cal I}(x^{t}_{1}) + \delta \right] $ &   $\eta_3^{t+1}= - \sigma   {D\cal I}(x^{t}_{1}) \eta_1^t + \left[ \beta  -4 \sigma  \right]  \sigma {D\cal I}(x^{t}_{3}) \eta_3^t$ \\ 
$x_1=x_2 \quad \& \quad x_3=x_4$ & $x^{t+1}_{4}=\left[ (\beta- 3 \sigma) {\cal I}(x^{t}_{4}) + 3 \sigma {\cal I}(x^{t}_{1}) + \delta \right] $ & \\
 \hline
    \multirow{3}{*} {A3: } 
&  $x^{t+1}_{1}=\left[ (\beta- 2 \sigma) {\cal I}(x^{t}_{1}) + \sigma {\cal I}(x^{t}_{2})+ \sigma {\cal I}(x^{t}_{5}) + \delta \right]$   & $\eta^{t+1}= \left[ \beta +\lambda \sigma  \right] {D\cal I}(x^{t}_{1}) \eta^t$, $\lambda=-3$  \\
 & $x^{t+1}_{2}=\left[ (\beta- 2 \sigma) {\cal I}(x^{t}_{2}) + \sigma {\cal I}(x^{t}_{1})  + \sigma {\cal I}(x^{t}_{5}) + \delta \right]$ &   \\ 
$x_1=x_3 \quad \& \quad x_2=x_4$ & $x^{t+1}_{5}=\left[ (\beta- 4 \sigma) {\cal I}(x^{t}_{5}) + 2 \sigma {\cal I}(x^{t}_{1})  + 2 \sigma {\cal I}(x^{t}_{2})+ \delta \right] $ & \\
  \hline
    \multirow{4}{*} {A4: $x_1=x_3$} 
&  $ x^{t+1}_{1}=\left[ (\beta- 3 \sigma) {\cal I}(x^{t}_{1}) + \sigma {\cal I}(x^{t}_{2})+   \sigma {\cal I}(x^{t}_{4}) + \sigma {\cal I}(x^{t}_{5}) + \delta \right] $   & $\eta^{t+1}= \left[ \beta +\lambda \sigma  \right] {D\cal I}(x^{t}_{1}) \eta^t$, $\lambda=-3$  \\
 & $x^{t+1}_{2}=\left[ (\beta- 3 \sigma) {\cal I}(x^{t}_{2}) + 2 \sigma {\cal I}(x^{t}_{1})  + \sigma {\cal I}(x^{t}_{5}) + \delta \right] $ &   \\ 
 & $x^{t+1}_{4}=\left[ (\beta- 3 \sigma) {\cal I}(x^{t}_{4}) + 2 \sigma {\cal I}(x^{t}_{1})  +  \sigma {\cal I}(x^{t}_{5})+ \delta \right] $ & \\
 & $ x^{t+1}_{5}=\left[ (\beta- 4 \sigma) {\cal I}(x^{t}_{5}) + 2 \sigma {\cal I}(x^{t}_{1})  +  \sigma {\cal I}(x^{t}_{2}) +  \sigma {\cal I}(x^{t}_{4}) + \delta \right] $ & \\
 \hline
 L1: $x_1=x_2=...=x_5$ & $x^{t+1}_{1}=\left[ \beta {\cal I}(x^{t}_{1})  + \delta \right]$   & $\eta^{t+1}= \left[ \beta  + \sigma \lambda  \right] {D\cal I}(x^{t}_{1}) \eta^t$, $\lambda=-3,-5$ \\ \hline
   \multirow{2}{*} {L3:} & $x^{t+1}_{1}=\left[ (\beta- 2 \sigma) {\cal I}(x^{t}_{1}) + 2 \sigma {\cal I}(x^{t}_{2}) + \delta \right] $   & $\eta^{t+1}_1= \left[ \beta  + \sigma \lambda  \right] {D\cal I}(x^{t}_{1}) \eta_1^t$, $\lambda=-3,-5$ \\
 $x_1=x_3=x_5 \&  x_2=x_4$ & $x^{t+1}_{2}=\left[ (\beta- 3 \sigma) {\cal I}(x^{t}_{2}) + 3 \sigma {\cal I}(x^{t}_{1}) + \delta \right] $ &  $\eta^{t+1}_2= \left[ \beta  + \sigma \lambda  \right] {D\cal I}(x^{t}_{2}) \eta_2^t$, $\lambda=-3$ \\
 \hline
 \multirow{3}{*} {L4: $x_1=x_3=x_5$} 
&  $x^{t+1}_{1}=\left[ (\beta- 2 \sigma) {\cal I}(x^{t}_{1}) + \sigma {\cal I}(x^{t}_{2})+ \sigma {\cal I}(x^{t}_{4}) + \delta \right] $   & $\eta^{t+1}= \left[ \beta  + \sigma \lambda  \right] {D\cal I}(x^{t}_{1}) \eta^t$, $\lambda=-3,-5$ \\
 & $x^{t+1}_{2}=\left[ (\beta- 3 \sigma) {\cal I}(x^{t}_{2}) + 3 \sigma {\cal I}(x^{t}_{1}) + \delta \right] $ &   \\ 
 & $x^{t+1}_{4}=\left[ (\beta- 3 \sigma) {\cal I}(x^{t}_{4}) + 3 \sigma {\cal I}(x^{t}_{1}) + \delta \right] $ & \\
 \hline
\end{tabular}
\end{center}
\caption{Equations used to evaluate stability of all the allowed CS patterns for the system described by Eq.\ \eqref{ExperimenalEquation} and coupling matrix  \eqref{eq:L}.}
\end{table}

As we showed in \cite{NC} we can first block diagonalize the coupling matrix $L$ using the irreducible representations (IRR) of ${\cal G'}$, which yields the transformed coupling matrix $L'$ for A3 of Fig.~\ref{fig:5node},

\begin{equation}\label{eq:L'}
L'=\begin{pmatrix}
   -4 & -\sqrt{2} & \sqrt{2}  & 0  & 0 \\
 -\sqrt{2} & -3 & -2  & 0  & 0 \\
 \sqrt{2} & -2 & -3  & 0  & 0 \\
  0  & 0  & 0 & -3  & 0 \\
  0  & 0  & 0  & 0 & -3 \\
  
 \end{pmatrix}
\end{equation} 

The upper-left block represents the variations on the synchronization manifold. It is $3 \times 3$ because there are three different trajectories in A3 (three clusters). The lower-right block represents the variations in the transverse manifold. These are the perturbations that take the system out of synchrony so it is these whose stability we want to calculate. Suppose we merge the center node (5) with nodes 1 and 3 to form the first merged state shown in L3. Geometrically, what must happen is that the dimension of the synchronization manifold  must decrease by one (from 3 to 2) and the transverse manifold must increase by 1  (from 2  to 3).  

\begin{figure}[!t]
\centering
\includegraphics[scale=2.5]{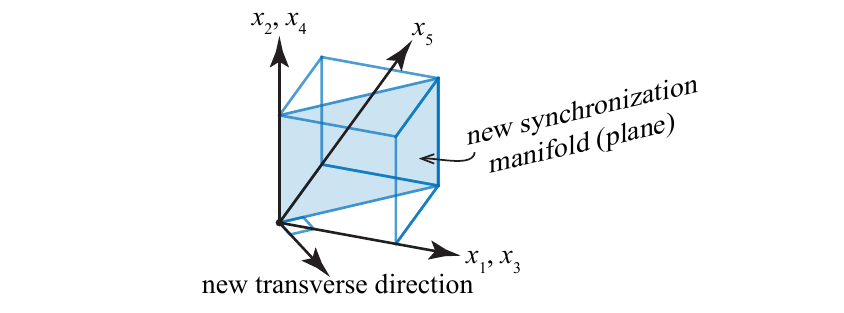}
\caption{Reduction of the dimension of the three-dimensional synchronization manifold. This shows schematically how the merging of clusters (1,3) and (5) produces a new synchronization direction in the  (1,3) and (5) plane of the synchronization manifold along with a new transverse direction orthogonal to the new synchronization direction.  \label{fig:syncman} }
\end{figure}

To obtain the new coordinates on the synchronization manifold, we note that basis vectors of a cluster on the synchronization manifold have a 1 in the position of each node that belongs to the cluster. For example, in A3 the cluster (1,3) will have a (unit) vector of the form (1,0,1,0,0) and (5) will have the (unit) vector of the form (0,0,0,0,1).  The merged cluster (1,3,5) will have the synchronization manifold vector which is their sum, (1,0,1,0,1).  The transformation of this new synchronization vector to the IRR coordinates of $L'$ provides the new synchronization direction and its orthogonal complement provides the new transverse direction.  We use these two new vectors to transform the $2 \times 2$ sub-block associated with the (1,3) and (5) clusters in the $3 \times 3$ synchronization block to reduce the $3 \times 3$ synchronization manifold and increase the transverse manifold.  This results in the final variational equation for the L3 case,


\begin{eqnarray}\label{eq:2}\nonumber 
\dot{{\pmb \eta}} &= & \sum_{m=1}^{M} \left[J^{(m)} \otimes D{\bf F}({\bf s}_m) + \sigma L''J^{(m)} \otimes D{\bf H}({\bf s}_m) \right] {\pmb \eta},
\end{eqnarray}

\noindent
where we have linearized about the new synchronized merged cluster states $\{{\bf s}_1,...,{\bf s}_M\}$, ${\pmb \eta}$ is the vector of variations of all nodes transformed to the merged coordinates as above, $D{\bf F}$ and $D{\bf H}$ are the Jacobians of the nodes' vector field and coupling function, respectively, $J^{(m)}$ are the transformed $E^{(m)}$ and,

\begin{equation}\label{eq:L''}
L''=\begin{pmatrix}
   -3 & \sqrt{6}  & 0  & 0  & 0 \\
 \sqrt{6} & -2  & 0  & 0  & 0 \\
  0  & 0 & -5  & 0  & 0 \\
  0  & 0  & 0 & -3  & 0 \\
  0  & 0  & 0  & 0 & -3 \\
  
 \end{pmatrix},
\end{equation}

\noindent
In Eq.(\ref{eq:L''}) the new synchronization block (in the upper-left-hand corner) represents the new clusters (1,3,5) and (2,4) and the new transverse direction is associated with the new diagonal value $-5$.  Obviously, this can be generalized to more complex cluster mergings. An example of cluster merging is shown geometrically in Fig.~\ref{fig:syncman}. We remark that this method can be used to analyze stability of any dynamically valid CS pattern, for which knowledge of the block-diagonalized matrix $L'$ is available and it applies to both  cases that the connectivity is given by an adjacency matrix or by a Laplacian matrix. The generalization of the above procedure simply uses the synchronization vectors for all the new clusters as the rows of a matrix for which all other components are zero. An application of a singular value decomposition then gives a basis for the synchronization block (the original synchronization vectors) and a basis for the orthogonal compliment which represents all the transverse directions. In some cases the latter can be simplified by a block diagonalization when all members of the block are in the same merged cluster.  This makes it possible to automate this block-diagonalization  to evaluate  stability of all the CS patterns that can emerge in a given network topology.

We show symmetry-breaking and the existence of Laplacian clusters using the experimental system described in  detail in \cite{Hagerstrom2012,NC}. This system employs a spatial light modulator (SLM) and a camera in a feedback configuration. The camera has a focal plane array (FPA) of 320 $\times$ 256 pixels, and an area of 8 $\times$ 6.4 mm$^2$. The SLM has a resolution of 512 $\times$ 512 pixels, and an active area of 7.68 $\times$ 7.68 mm$^2$. A light emitting diode with a wavelength of 1550 nm is used to illuminate the modulator. The light passes through a polarizing beam-splitter, and a quarter-wave plate (QWP), so that circularly polarized light is incident on the SLM. The SLM imparts a programmable spatially-varying phase shift $x$ between the two polarization components of the reflected light. The reflected light passes through the QWP and polarizer, and is imaged onto a 256 $\times$ 256 pixel square region of the camera's FPA. The relationship between the phase shift $x$ applied by the SLM and the normalized intensity ${\cal I}$ recorded by the camera is ${\cal I}(x)=\left( 1-\cos x \right)/2$.

Each oscillator corresponds to a square patch of 16  $\times$ 16 pixels on the SLM, which is imaged onto an 8  $\times$ 8 pixel region of the camera's FPA.
Using a computer, the phase shift of the $i$th region, $x_i$, is iteratively updated based on the intensity measured by the camera according to:

\begin{equation}
x^{t+1}_i=\left[ \beta {\cal I}(x^{t}_i) + \sigma \sum_j L_{ij} {\cal I}(x^{t}_j) + \delta \right] \text{ mod } 2\pi \label{ExperimenalEquation}
\end{equation}

\noindent
where $\sigma$ is the coupling strength, which we will vary.  Eq.~(\ref{ExperimenalEquation}) is a map version of Eq.~(\ref{eq:1}) and can show fixed point, periodic or chaotic dynamics depending on the values of the parameters. Here, $\delta=0.525$, $\beta=1.45 \pi$, and $\sigma$ was decreased from $\pi$ to $0$. 

\begin{figure}[h]
\centering
\includegraphics[scale=1]{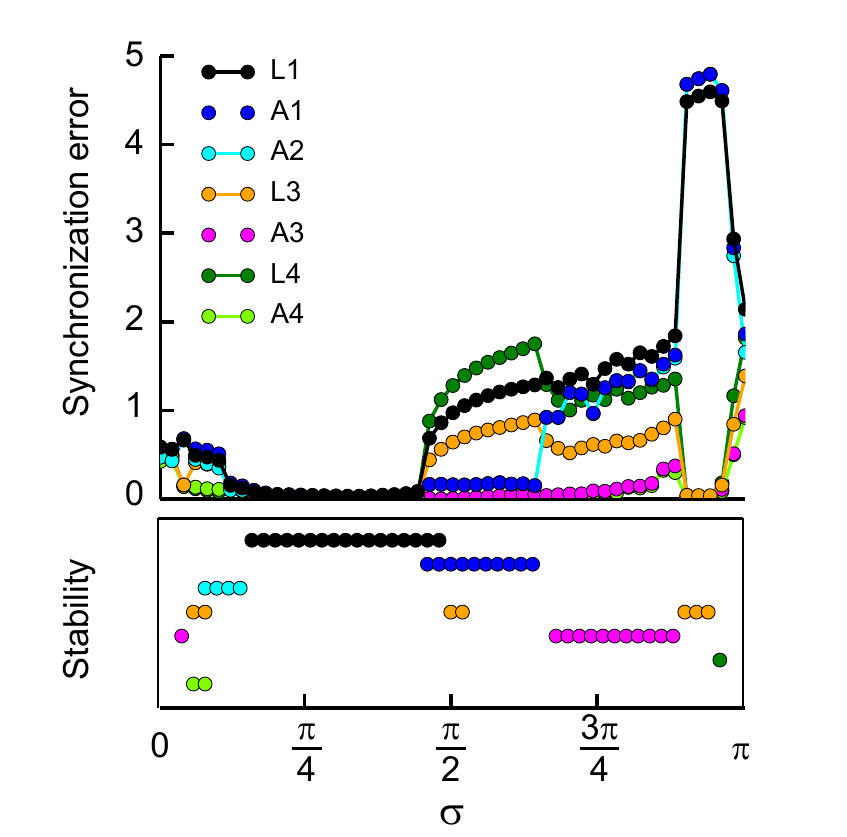}
\caption{The figure shows the experimental synchronization error for each synchronization pattern as a function of the parameter $\sigma$. Underneath we plot the results of our stability analysis applied to each one of the cluster synchronization patterns, where a colored dot labels the values of sigma for which the corresponding pattern is stable. \label{fig:syncError}}
\end{figure}

Fig. \ref{fig:syncError} shows the experimental synchronization error for each synchronization pattern shown in Fig. \ref{fig:5node} as a function of $\sigma$. The phases $x_i$ were not reinitialized when $\sigma$ was updated. The synchronization error for each pattern is computed as $\left \langle |x_i(t)- x^{cl}_i(t)| \right \rangle$, where the symbol $\left \langle \bullet \right \rangle$ indicates an average both in time and over the nodes $i$ in a cluster and the clusters in a pattern. $x^{cl}_i$ is the average state for the nodes in the cluster to which node $i$ belongs. In the lower panel of Fig. \ref{fig:syncError}, we plot the results of our stability analysis for each one of the CS patterns, where a colored dot labels the values of sigma for which the corresponding pattern is stable. In particular, a CS pattern was indicated to be stable when (i) all the numerically computed maximum Lyapunov exponents corresponding to the transverse blocks were found to be negative and (ii) the synchronous pattern was \emph{asymptotically valid}, that is the CS pattern was observed after integrating its equations for a long time. The equations that were used to run these stability calculations are shown in Table I.

In general, when two or more clusters merge into one, there are two independent effects on stability, as can be seen from the structure of the block-diagonalized matrix $L''$: the first one is that the dimension of the synchronization block decreases, which determines the motion in the synchronization manifold; the second one is that new transverse blocks appear, with the other (pre-existing) transverse blocks remaining the same.  As a consequence, since  each one of these transverse blocks needs to be individually tested for stability, we expect that as the number of transversal blocks increases, the  range of stability decreases. This is confirmed by our experimental results plotted in Fig.  \ref{fig:syncError}, showing that the $\sigma$-range of stability becomes smaller for CS patterns that are characterized by higher symmetry.
Exceptions to this rule are possible as the motion in the synchronization manifold (on which the transverse Lyapunov exponents depend) 
may also affect stability in ways that cannot be predicted by the analysis of the transverse blocks only.

\begin{figure}[t!]
\centering
\includegraphics[scale=1]{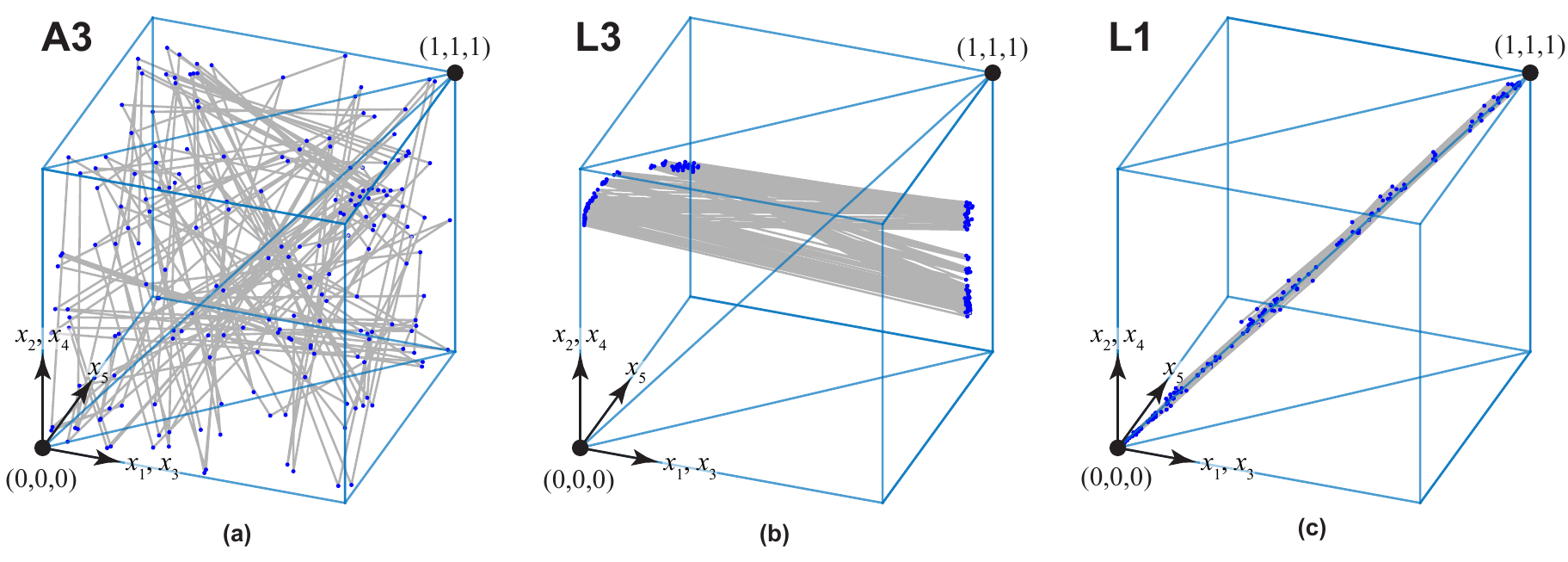}
\caption{Experimental phase space plots with lines connecting
 successive iterates for  a) three clusters (A3), b)  two clusters (L3) 
 and c) one cluster (L1).   \label{EXP_SYNCMAN} }
\end{figure}
\begin{figure}[h]
\centering
\includegraphics[scale=1,trim=0.5cm 11cm 0.5cm 0cm,]{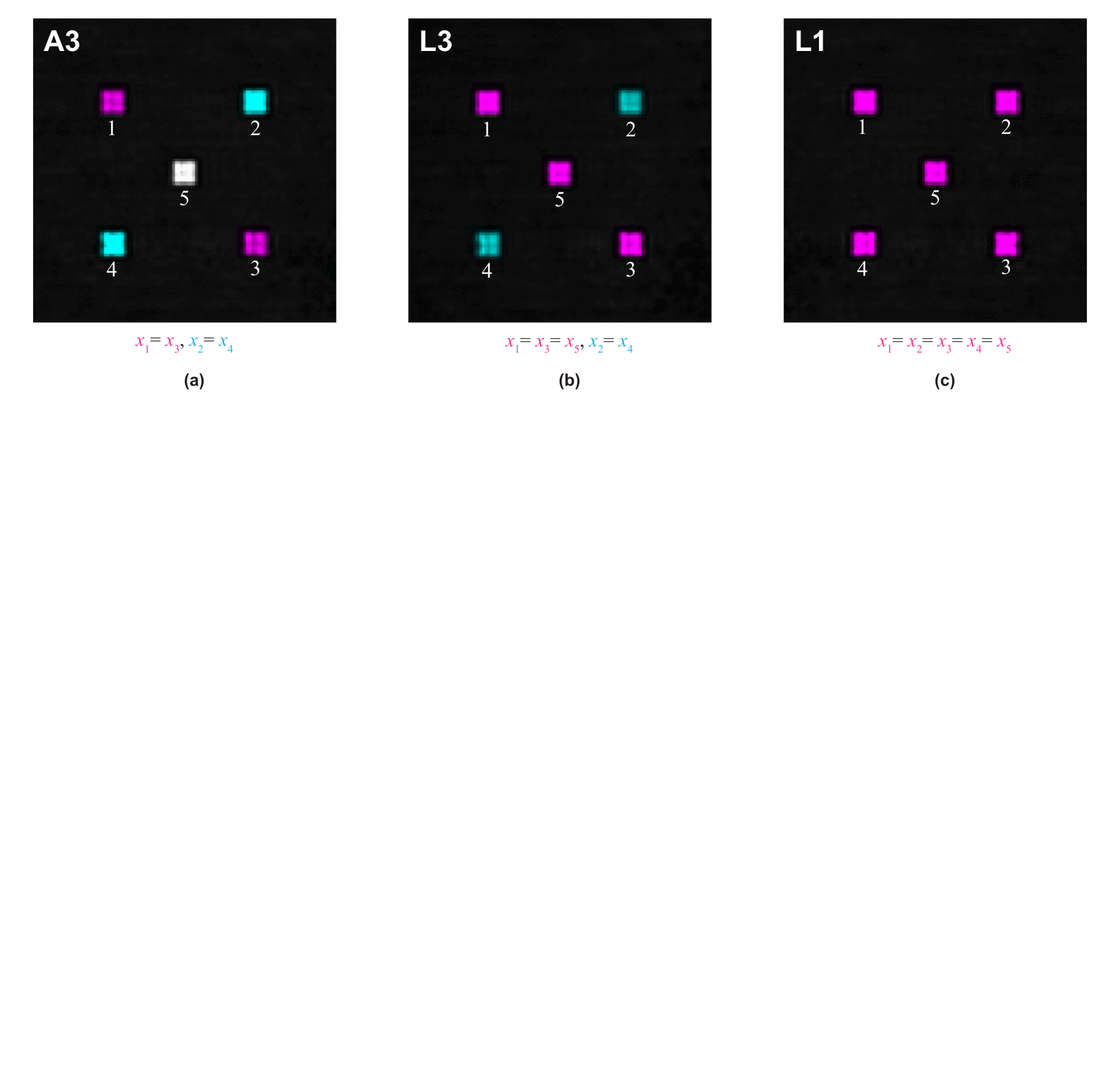}
\caption{Snapshots of the experimental dynamics for the cases of  a) three clusters (A3), b)  two clusters (L3) 
 and c) one cluster (L1).   \label{EXP_SYNCMAN2} }
\end{figure}

Fig. \ref{EXP_SYNCMAN} shows the dynamics on the synchronization manifold for the symmetry pattern A3 and the two merged patterns, L3 and L1. In each transition A3 $\rightarrow$ L3 $\rightarrow$ L1 the dimension of the synchronization manifold decreases by 1 (from 3 to 2 to 1) and the transverse manifold increases by 1 (from 2 to 3 to 4).  Fig. \ref{EXP_SYNCMAN2} shows three snapshots of the experimental dynamics for each one of the patterns A3,L3, and L1. Videos of the synchronization manifolds and patterns of Fig. \ref{EXP_SYNCMAN2} are available in the supplement.

\section{Conclusion}
 
We have studied the emergence of cluster synchronization in networks of coupled oscillators. First, we show how to obtain all of the possible dynamically valid CS patterns for an arbitrary network topology.
The methods we illustrated {above} are general and will apply to nodes with any type of dynamics (ODEs, maps, etc.).  This method also extends to directed networks, weighted networks, and labeled nodes (to represent different dynamics on each node), all of which are handled by the software package \cite{Stein}.  An important point is that these techniques work for any subgroups of the cluster groups which includes the trivial subgroups \cite{NC} (only the identity in each).  Hence, we can approach merging as a bottom up process and analyze any arbitrary merging of nodes into a cluster to determine if the dynamics allows a synchronized state and if it is stable in some parameter range.  This means all possible clusters can be analyzed using our approach.  We note that combining clusters as a ``top down" approach would provide clusters that most likely are not from symmetries whereas bottom up would be a process that would include clusters easily obtained from symmetries, although it would be useful for cases where one has particular clusters in mind to analyze.

 Our main result is a technique to evaluate stability of all the dynamically valid CS patterns for both networks for which the connectivity is given by an adjacency matrix and by a Laplacian matrix. We predict that the range of stability typically becomes smaller for CS patterns that are characterized by higher symmetry, which is confirmed in our experimental system.


We acknowledge help with computational group theory from Prof. David Joyner of the US Naval Academy and information and computer code from Ben D. MacArthur and Rub\'en J. S\'anchez-Garc\'ia both of the University of Southampton. This work was funded by the Office of Naval Research. F.S. was supported by NSF grant CMMI-1400193.

\bibliographystyle{apsrev4-1} 
\bibliography{NatCommBib,biblioRidond2}

\begin{thebibliography}{38}%
\makeatletter
\providecommand \@ifxundefined [1]{%
 \@ifx{#1\undefined}
}%
\providecommand \@ifnum [1]{%
 \ifnum #1\expandafter \@firstoftwo
 \else \expandafter \@secondoftwo
 \fi
}%
\providecommand \@ifx [1]{%
 \ifx #1\expandafter \@firstoftwo
 \else \expandafter \@secondoftwo
 \fi
}%
\providecommand \natexlab [1]{#1}%
\providecommand \enquote  [1]{``#1''}%
\providecommand \bibnamefont  [1]{#1}%
\providecommand \bibfnamefont [1]{#1}%
\providecommand \citenamefont [1]{#1}%
\providecommand \href@noop [0]{\@secondoftwo}%
\providecommand \href [0]{\begingroup \@sanitize@url \@href}%
\providecommand \@href[1]{\@@startlink{#1}\@@href}%
\providecommand \@@href[1]{\endgroup#1\@@endlink}%
\providecommand \@sanitize@url [0]{\catcode `\\12\catcode `\$12\catcode
  `\&12\catcode `\#12\catcode `\^12\catcode `\_12\catcode `\%12\relax}%
\providecommand \@@startlink[1]{}%
\providecommand \@@endlink[0]{}%
\providecommand \url  [0]{\begingroup\@sanitize@url \@url }%
\providecommand \@url [1]{\endgroup\@href {#1}{\urlprefix }}%
\providecommand \urlprefix  [0]{URL }%
\providecommand \Eprint [0]{\href }%
\providecommand \doibase [0]{http://dx.doi.org/}%
\providecommand \selectlanguage [0]{\@gobble}%
\providecommand \bibinfo  [0]{\@secondoftwo}%
\providecommand \bibfield  [0]{\@secondoftwo}%
\providecommand \translation [1]{[#1]}%
\providecommand \BibitemOpen [0]{}%
\providecommand \bibitemStop [0]{}%
\providecommand \bibitemNoStop [0]{.\EOS\space}%
\providecommand \EOS [0]{\spacefactor3000\relax}%
\providecommand \BibitemShut  [1]{\csname bibitem#1\endcsname}%
\let\auto@bib@innerbib\@empty
\bibitem [{\citenamefont {DeMarco}\ and\ \citenamefont
  {Wassner}(1995)}]{DeMarco1995}%
  \BibitemOpen
  \bibfield  {author} {\bibinfo {author} {\bibfnamefont {C.~L.}\ \bibnamefont
  {DeMarco}}\ and\ \bibinfo {author} {\bibfnamefont {J.}~\bibnamefont
  {Wassner}},\ }in\ \href@noop {} {\emph {\bibinfo {booktitle} {Control
  Applications, 1995, Proceedings of the 4th IEEE Conference on}}}\ (\bibinfo
  {organization} {IEEE},\ \bibinfo {year} {1995})\ pp.\ \bibinfo {pages}
  {611--617}\BibitemShut {NoStop}%
\bibitem [{\citenamefont {Motter}\ \emph {et~al.}(2013)\citenamefont {Motter},
  \citenamefont {Myers}, \citenamefont {Anghel},\ and\ \citenamefont
  {Nishikawa}}]{Motter2013}%
  \BibitemOpen
  \bibfield  {author} {\bibinfo {author} {\bibfnamefont {A.~E.}\ \bibnamefont
  {Motter}}, \bibinfo {author} {\bibfnamefont {S.~A.}\ \bibnamefont {Myers}},
  \bibinfo {author} {\bibfnamefont {M.}~\bibnamefont {Anghel}}, \ and\ \bibinfo
  {author} {\bibfnamefont {T.}~\bibnamefont {Nishikawa}},\ }\href@noop {}
  {\bibfield  {journal} {\bibinfo  {journal} {Nat. Phys.}\ }\textbf {\bibinfo
  {volume} {9}},\ \bibinfo {pages} {191} (\bibinfo {year} {2013})}\BibitemShut
  {NoStop}%
\bibitem [{\citenamefont {Abdollahy}\ \emph {et~al.}(2012)\citenamefont
  {Abdollahy}, \citenamefont {Lavrova}, \citenamefont {Mammoli}, \citenamefont
  {Willard},\ and\ \citenamefont {Arellano}}]{abdollahy2012pnm}%
  \BibitemOpen
  \bibfield  {author} {\bibinfo {author} {\bibfnamefont {S.}~\bibnamefont
  {Abdollahy}}, \bibinfo {author} {\bibfnamefont {O.}~\bibnamefont {Lavrova}},
  \bibinfo {author} {\bibfnamefont {A.}~\bibnamefont {Mammoli}}, \bibinfo
  {author} {\bibfnamefont {S.}~\bibnamefont {Willard}}, \ and\ \bibinfo
  {author} {\bibfnamefont {B.}~\bibnamefont {Arellano}},\ }in\ \href@noop {}
  {\emph {\bibinfo {booktitle} {Innovative Smart Grid Technologies (ISGT), 2012
  IEEE PES}}}\ (\bibinfo {organization} {IEEE},\ \bibinfo {year} {2012})\ pp.\
  \bibinfo {pages} {1--6}\BibitemShut {NoStop}%
\bibitem [{\citenamefont {Boccaletti}\ \emph {et~al.}(2002)\citenamefont
  {Boccaletti}, \citenamefont {Kurths}, \citenamefont {Osipov}, \citenamefont
  {Valladares},\ and\ \citenamefont {Zhou}}]{Boccaletti2002}%
  \BibitemOpen
  \bibfield  {author} {\bibinfo {author} {\bibfnamefont {S.}~\bibnamefont
  {Boccaletti}}, \bibinfo {author} {\bibfnamefont {J.}~\bibnamefont {Kurths}},
  \bibinfo {author} {\bibfnamefont {G.}~\bibnamefont {Osipov}}, \bibinfo
  {author} {\bibfnamefont {D.}~\bibnamefont {Valladares}}, \ and\ \bibinfo
  {author} {\bibfnamefont {C.}~\bibnamefont {Zhou}},\ }\href@noop {} {\bibfield
   {journal} {\bibinfo  {journal} {Physics reports}\ }\textbf {\bibinfo
  {volume} {366}},\ \bibinfo {pages} {1} (\bibinfo {year} {2002})}\BibitemShut
  {NoStop}%
\bibitem [{\citenamefont {Pecora}\ and\ \citenamefont
  {Carroll}(1998{\natexlab{a}})}]{Pecora1998master}%
  \BibitemOpen
  \bibfield  {author} {\bibinfo {author} {\bibfnamefont {L.~M.}\ \bibnamefont
  {Pecora}}\ and\ \bibinfo {author} {\bibfnamefont {T.~L.}\ \bibnamefont
  {Carroll}},\ }\href@noop {} {\bibfield  {journal} {\bibinfo  {journal} {Phys.
  Rev. Lett.}\ }\textbf {\bibinfo {volume} {80}},\ \bibinfo {pages} {2109}
  (\bibinfo {year} {1998}{\natexlab{a}})}\BibitemShut {NoStop}%
\bibitem [{\citenamefont {Pikovsky}\ \emph {et~al.}(2003)\citenamefont
  {Pikovsky}, \citenamefont {Rosenblum},\ and\ \citenamefont
  {Kurths}}]{PikovskyBOOK2003}%
  \BibitemOpen
  \bibfield  {author} {\bibinfo {author} {\bibfnamefont {A.}~\bibnamefont
  {Pikovsky}}, \bibinfo {author} {\bibfnamefont {M.}~\bibnamefont {Rosenblum}},
  \ and\ \bibinfo {author} {\bibfnamefont {J.}~\bibnamefont {Kurths}},\
  }\href@noop {} {\emph {\bibinfo {title} {Synchronization: a universal concept
  in nonlinear sciences}}},\ Vol.~\bibinfo {volume} {12}\ (\bibinfo
  {publisher} {Cambridge university press},\ \bibinfo {year}
  {2003})\BibitemShut {NoStop}%
\bibitem [{\citenamefont {Zhou}\ and\ \citenamefont {Kurths}(2006)}]{Zhou2006}%
  \BibitemOpen
  \bibfield  {author} {\bibinfo {author} {\bibfnamefont {C.}~\bibnamefont
  {Zhou}}\ and\ \bibinfo {author} {\bibfnamefont {J.}~\bibnamefont {Kurths}},\
  }\href@noop {} {\bibfield  {journal} {\bibinfo  {journal} {Chaos}\ }\textbf
  {\bibinfo {volume} {16}},\ \bibinfo {pages} {015104} (\bibinfo {year}
  {2006})}\BibitemShut {NoStop}%
\bibitem [{\citenamefont {A-L.Do}\ \emph {et~al.}(2012)\citenamefont {A-L.Do},
  \citenamefont {Hoefener},\ and\ \citenamefont {Gross}}]{Do2012}%
  \BibitemOpen
  \bibfield  {author} {\bibinfo {author} {\bibnamefont {A-L.Do}}, \bibinfo
  {author} {\bibfnamefont {J.}~\bibnamefont {Hoefener}}, \ and\ \bibinfo
  {author} {\bibfnamefont {T.}~\bibnamefont {Gross}},\ }\href@noop {}
  {\bibfield  {journal} {\bibinfo  {journal} {New Journal of Physics}\ }\textbf
  {\bibinfo {volume} {14}},\ \bibinfo {pages} {115022} (\bibinfo {year}
  {2012})}\BibitemShut {NoStop}%
\bibitem [{\citenamefont {Dahms}\ \emph {et~al.}(2012)\citenamefont {Dahms},
  \citenamefont {Lehnert},\ and\ \citenamefont {Sch\"{o}ll}}]{Dahms2012}%
  \BibitemOpen
  \bibfield  {author} {\bibinfo {author} {\bibfnamefont {T.}~\bibnamefont
  {Dahms}}, \bibinfo {author} {\bibfnamefont {J.}~\bibnamefont {Lehnert}}, \
  and\ \bibinfo {author} {\bibfnamefont {E.}~\bibnamefont {Sch\"{o}ll}},\
  }\href@noop {} {\bibfield  {journal} {\bibinfo  {journal} {Physical Review
  E}\ }\textbf {\bibinfo {volume} {86}},\ \bibinfo {pages} {016202} (\bibinfo
  {year} {2012})}\BibitemShut {NoStop}%
\bibitem [{\citenamefont {Fu}\ \emph {et~al.}(2013)\citenamefont {Fu},
  \citenamefont {Deng}, \citenamefont {Huang},\ and\ \citenamefont
  {Wang}}]{Fu2013}%
  \BibitemOpen
  \bibfield  {author} {\bibinfo {author} {\bibfnamefont {C.}~\bibnamefont
  {Fu}}, \bibinfo {author} {\bibfnamefont {Z.}~\bibnamefont {Deng}}, \bibinfo
  {author} {\bibfnamefont {L.}~\bibnamefont {Huang}}, \ and\ \bibinfo {author}
  {\bibfnamefont {X.}~\bibnamefont {Wang}},\ }\href@noop {} {\bibfield
  {journal} {\bibinfo  {journal} {Physical Review E}\ }\textbf {\bibinfo
  {volume} {87}},\ \bibinfo {pages} {032909} (\bibinfo {year}
  {2013})}\BibitemShut {NoStop}%
\bibitem [{\citenamefont {Kanter}\ \emph {et~al.}(2011)\citenamefont {Kanter},
  \citenamefont {Zigzag}, \citenamefont {Englert}, \citenamefont {Geissler},\
  and\ \citenamefont {Kinzel}}]{Kanter2011}%
  \BibitemOpen
  \bibfield  {author} {\bibinfo {author} {\bibfnamefont {I.}~\bibnamefont
  {Kanter}}, \bibinfo {author} {\bibfnamefont {M.}~\bibnamefont {Zigzag}},
  \bibinfo {author} {\bibfnamefont {A.}~\bibnamefont {Englert}}, \bibinfo
  {author} {\bibfnamefont {F.}~\bibnamefont {Geissler}}, \ and\ \bibinfo
  {author} {\bibfnamefont {W.}~\bibnamefont {Kinzel}},\ }\href@noop {}
  {\bibfield  {journal} {\bibinfo  {journal} {EPL}\ }\textbf {\bibinfo {volume}
  {93}},\ \bibinfo {pages} {6003} (\bibinfo {year} {2011})}\BibitemShut
  {NoStop}%
\bibitem [{\citenamefont {Rosin}\ \emph {et~al.}(2013)\citenamefont {Rosin},
  \citenamefont {Rontani}, \citenamefont {Gauthier},\ and\ \citenamefont
  {Sch\"{o}ll}}]{Rosin2013}%
  \BibitemOpen
  \bibfield  {author} {\bibinfo {author} {\bibfnamefont {D.~P.}\ \bibnamefont
  {Rosin}}, \bibinfo {author} {\bibfnamefont {D.}~\bibnamefont {Rontani}},
  \bibinfo {author} {\bibfnamefont {D.}~\bibnamefont {Gauthier}}, \ and\
  \bibinfo {author} {\bibfnamefont {E.}~\bibnamefont {Sch\"{o}ll}},\
  }\href@noop {} {\bibfield  {journal} {\bibinfo  {journal} {Phys. Rev. Lett.}\
  }\textbf {\bibinfo {volume} {110}},\ \bibinfo {pages} {104102} (\bibinfo
  {year} {2013})}\BibitemShut {NoStop}%
\bibitem [{\citenamefont {Sorrentino}\ and\ \citenamefont
  {Ott}(2007)}]{Sorrentino2007}%
  \BibitemOpen
  \bibfield  {author} {\bibinfo {author} {\bibfnamefont {F.}~\bibnamefont
  {Sorrentino}}\ and\ \bibinfo {author} {\bibfnamefont {E.}~\bibnamefont
  {Ott}},\ }\href@noop {} {\bibfield  {journal} {\bibinfo  {journal} {Physical
  Review E}\ }\textbf {\bibinfo {volume} {76}},\ \bibinfo {pages} {056114}
  (\bibinfo {year} {2007})}\BibitemShut {NoStop}%
\bibitem [{\citenamefont {Williams}\ \emph {et~al.}(2013)\citenamefont
  {Williams}, \citenamefont {Murphy}, \citenamefont {Roy}, \citenamefont
  {Sorrentino}, \citenamefont {Dahms},\ and\ \citenamefont
  {Sch{\"o}ll}}]{Williams2013}%
  \BibitemOpen
  \bibfield  {author} {\bibinfo {author} {\bibfnamefont {C.~R.}\ \bibnamefont
  {Williams}}, \bibinfo {author} {\bibfnamefont {T.~E.}\ \bibnamefont
  {Murphy}}, \bibinfo {author} {\bibfnamefont {R.}~\bibnamefont {Roy}},
  \bibinfo {author} {\bibfnamefont {F.}~\bibnamefont {Sorrentino}}, \bibinfo
  {author} {\bibfnamefont {T.}~\bibnamefont {Dahms}}, \ and\ \bibinfo {author}
  {\bibfnamefont {E.}~\bibnamefont {Sch{\"o}ll}},\ }\href@noop {} {\bibfield
  {journal} {\bibinfo  {journal} {Phys. Rev. Lett.}\ }\textbf {\bibinfo
  {volume} {110}},\ \bibinfo {pages} {064104} (\bibinfo {year}
  {2013})}\BibitemShut {NoStop}%
\bibitem [{\citenamefont {D'Huys}\ \emph {et~al.}(2008)\citenamefont {D'Huys},
  \citenamefont {Vicente}, \citenamefont {Erneux}, \citenamefont {Danckaert},\
  and\ \citenamefont {Fischer}}]{D'Huys2008}%
  \BibitemOpen
  \bibfield  {author} {\bibinfo {author} {\bibfnamefont {O.}~\bibnamefont
  {D'Huys}}, \bibinfo {author} {\bibfnamefont {R.}~\bibnamefont {Vicente}},
  \bibinfo {author} {\bibfnamefont {T.}~\bibnamefont {Erneux}}, \bibinfo
  {author} {\bibfnamefont {J.}~\bibnamefont {Danckaert}}, \ and\ \bibinfo
  {author} {\bibfnamefont {I.}~\bibnamefont {Fischer}},\ }\href@noop {}
  {\bibfield  {journal} {\bibinfo  {journal} {Chaos}\ }\textbf {\bibinfo
  {volume} {18}},\ \bibinfo {pages} {037116} (\bibinfo {year}
  {2008})}\BibitemShut {NoStop}%
\bibitem [{\citenamefont {Orosz}(2014)}]{Gabor}%
  \BibitemOpen
  \bibfield  {author} {\bibinfo {author} {\bibfnamefont {G.}~\bibnamefont
  {Orosz}},\ }\href@noop {} {\bibfield  {journal} {\bibinfo  {journal} {SIAM
  Journal on Applied Dynamical Systems}\ }\textbf {\bibinfo {volume} {13}},\
  \bibinfo {pages} {1353} (\bibinfo {year} {2014})}\BibitemShut {NoStop}%
\bibitem [{\citenamefont {Belykh}\ \emph {et~al.}(2005)\citenamefont {Belykh},
  \citenamefont {de~Lange},\ and\ \citenamefont
  {Hasler}}]{belykh2005synchronization}%
  \BibitemOpen
  \bibfield  {author} {\bibinfo {author} {\bibfnamefont {I.}~\bibnamefont
  {Belykh}}, \bibinfo {author} {\bibfnamefont {E.}~\bibnamefont {de~Lange}}, \
  and\ \bibinfo {author} {\bibfnamefont {M.}~\bibnamefont {Hasler}},\
  }\href@noop {} {\bibfield  {journal} {\bibinfo  {journal} {Physical review
  letters}\ }\textbf {\bibinfo {volume} {94}},\ \bibinfo {pages} {188101}
  (\bibinfo {year} {2005})}\BibitemShut {NoStop}%
\bibitem [{\citenamefont {Cohen}\ \emph {et~al.}(2010)\citenamefont {Cohen},
  \citenamefont {Ravoori}, \citenamefont {Sorrentino}, \citenamefont {Murphy},
  \citenamefont {Ott},\ and\ \citenamefont {Roy}}]{cohen2010dynamic}%
  \BibitemOpen
  \bibfield  {author} {\bibinfo {author} {\bibfnamefont {A.~B.}\ \bibnamefont
  {Cohen}}, \bibinfo {author} {\bibfnamefont {B.}~\bibnamefont {Ravoori}},
  \bibinfo {author} {\bibfnamefont {F.}~\bibnamefont {Sorrentino}}, \bibinfo
  {author} {\bibfnamefont {T.~E.}\ \bibnamefont {Murphy}}, \bibinfo {author}
  {\bibfnamefont {E.}~\bibnamefont {Ott}}, \ and\ \bibinfo {author}
  {\bibfnamefont {R.}~\bibnamefont {Roy}},\ }\href@noop {} {\bibfield
  {journal} {\bibinfo  {journal} {Chaos: An Interdisciplinary Journal of
  Nonlinear Science}\ }\textbf {\bibinfo {volume} {20}},\ \bibinfo {pages}
  {043142} (\bibinfo {year} {2010})}\BibitemShut {NoStop}%
\bibitem [{\citenamefont {Belykh}\ \emph {et~al.}(2008)\citenamefont {Belykh},
  \citenamefont {Osipov}, \citenamefont {Petrov}, \citenamefont {Suykens},\
  and\ \citenamefont {Vandewalle}}]{Belykh2008}%
  \BibitemOpen
  \bibfield  {author} {\bibinfo {author} {\bibfnamefont {V.}~\bibnamefont
  {Belykh}}, \bibinfo {author} {\bibfnamefont {G.}~\bibnamefont {Osipov}},
  \bibinfo {author} {\bibfnamefont {V.}~\bibnamefont {Petrov}}, \bibinfo
  {author} {\bibfnamefont {J.}~\bibnamefont {Suykens}}, \ and\ \bibinfo
  {author} {\bibfnamefont {J.}~\bibnamefont {Vandewalle}},\ }\href@noop {}
  {\bibfield  {journal} {\bibinfo  {journal} {Chaos}\ }\textbf {\bibinfo
  {volume} {18}},\ \bibinfo {pages} {037106} (\bibinfo {year}
  {2008})}\BibitemShut {NoStop}%
\bibitem [{\citenamefont {Russo}\ and\ \citenamefont
  {Slotine}(2011)}]{Russo2011}%
  \BibitemOpen
  \bibfield  {author} {\bibinfo {author} {\bibfnamefont {G.}~\bibnamefont
  {Russo}}\ and\ \bibinfo {author} {\bibfnamefont {J.-J.~E.}\ \bibnamefont
  {Slotine}},\ }\href@noop {} {\bibfield  {journal} {\bibinfo  {journal}
  {Physical Review E}\ }\textbf {\bibinfo {volume} {84}},\ \bibinfo {pages}
  {041929} (\bibinfo {year} {2011})}\BibitemShut {NoStop}%
\bibitem [{\citenamefont {Golubitsky}\ and\ \citenamefont
  {Stewart}(2002)}]{GolubitskySewartBOOK}%
  \BibitemOpen
  \bibfield  {author} {\bibinfo {author} {\bibfnamefont {M.}~\bibnamefont
  {Golubitsky}}\ and\ \bibinfo {author} {\bibfnamefont {I.}~\bibnamefont
  {Stewart}},\ }\href@noop {} {\emph {\bibinfo {title} {The Symmetry
  Perspective: From Equilibrium to Chaos in Phase Space and Physical Space}}},\
  Vol.~\bibinfo {volume} {II}\ (\bibinfo  {publisher} {Berkh\"auser-Verlag},\
  \bibinfo {address} {Basel},\ \bibinfo {year} {2002})\BibitemShut {NoStop}%
\bibitem [{\citenamefont {Golubitsky}\ \emph {et~al.}(1985)\citenamefont
  {Golubitsky}, \citenamefont {Stewart},\ and\ \citenamefont
  {Schaeffer}}]{GolubitskyBOOKII}%
  \BibitemOpen
  \bibfield  {author} {\bibinfo {author} {\bibfnamefont {M.}~\bibnamefont
  {Golubitsky}}, \bibinfo {author} {\bibfnamefont {I.}~\bibnamefont {Stewart}},
  \ and\ \bibinfo {author} {\bibfnamefont {D.}~\bibnamefont {Schaeffer}},\
  }\href@noop {} {\emph {\bibinfo {title} {Singularities and groups in
  bifurcation theory}}},\ Vol.~\bibinfo {volume} {II}\ (\bibinfo  {publisher}
  {Springer-Verlag},\ \bibinfo {address} {New York, NY},\ \bibinfo {year}
  {1985})\BibitemShut {NoStop}%
\bibitem [{\citenamefont {Nicosia}\ \emph {et~al.}(2013)\citenamefont
  {Nicosia}, \citenamefont {Valencia}, \citenamefont {Chavez}, \citenamefont
  {D{\'\i}az-Guilera},\ and\ \citenamefont {Latora}}]{Nicosia2013}%
  \BibitemOpen
  \bibfield  {author} {\bibinfo {author} {\bibfnamefont {V.}~\bibnamefont
  {Nicosia}}, \bibinfo {author} {\bibfnamefont {M.}~\bibnamefont {Valencia}},
  \bibinfo {author} {\bibfnamefont {M.}~\bibnamefont {Chavez}}, \bibinfo
  {author} {\bibfnamefont {A.}~\bibnamefont {D{\'\i}az-Guilera}}, \ and\
  \bibinfo {author} {\bibfnamefont {V.}~\bibnamefont {Latora}},\ }\href@noop {}
  {\bibfield  {journal} {\bibinfo  {journal} {Phys. Rev. Lett.}\ }\textbf
  {\bibinfo {volume} {110}},\ \bibinfo {pages} {174102} (\bibinfo {year}
  {2013})}\BibitemShut {NoStop}%
\bibitem [{\citenamefont {Pecora}\ \emph {et~al.}(2014)\citenamefont {Pecora},
  \citenamefont {Sorrentino}, \citenamefont {Hagerstrom}, \citenamefont
  {Murphy},\ and\ \citenamefont {Roy}}]{NC}%
  \BibitemOpen
  \bibfield  {author} {\bibinfo {author} {\bibfnamefont {L.~M.}\ \bibnamefont
  {Pecora}}, \bibinfo {author} {\bibfnamefont {F.}~\bibnamefont {Sorrentino}},
  \bibinfo {author} {\bibfnamefont {A.~M.}\ \bibnamefont {Hagerstrom}},
  \bibinfo {author} {\bibfnamefont {T.~E.}\ \bibnamefont {Murphy}}, \ and\
  \bibinfo {author} {\bibfnamefont {R.}~\bibnamefont {Roy}},\ }\href@noop {}
  {\bibfield  {journal} {\bibinfo  {journal} {Nature Communications}\ }\textbf
  {\bibinfo {volume} {5}} (\bibinfo {year} {2014})}\BibitemShut {NoStop}%
\bibitem [{\citenamefont {Ravoori}\ \emph {et~al.}(2009)\citenamefont
  {Ravoori}, \citenamefont {Cohen}, \citenamefont {Setty}, \citenamefont
  {Sorrentino}, \citenamefont {Murphy}, \citenamefont {Ott},\ and\
  \citenamefont {Roy}}]{ravoori2009adaptive}%
  \BibitemOpen
  \bibfield  {author} {\bibinfo {author} {\bibfnamefont {B.}~\bibnamefont
  {Ravoori}}, \bibinfo {author} {\bibfnamefont {A.~B.}\ \bibnamefont {Cohen}},
  \bibinfo {author} {\bibfnamefont {A.~V.}\ \bibnamefont {Setty}}, \bibinfo
  {author} {\bibfnamefont {F.}~\bibnamefont {Sorrentino}}, \bibinfo {author}
  {\bibfnamefont {T.~E.}\ \bibnamefont {Murphy}}, \bibinfo {author}
  {\bibfnamefont {E.}~\bibnamefont {Ott}}, \ and\ \bibinfo {author}
  {\bibfnamefont {R.}~\bibnamefont {Roy}},\ }\href@noop {} {\bibfield
  {journal} {\bibinfo  {journal} {Physical Review E}\ }\textbf {\bibinfo
  {volume} {80}},\ \bibinfo {pages} {056205} (\bibinfo {year}
  {2009})}\BibitemShut {NoStop}%
\bibitem [{\citenamefont {Golubitsky}\ \emph {et~al.}(2005)\citenamefont
  {Golubitsky}, \citenamefont {Stewart},\ and\ \citenamefont
  {T{\"o}r{\"o}k}}]{golubitsky2005patterns}%
  \BibitemOpen
  \bibfield  {author} {\bibinfo {author} {\bibfnamefont {M.}~\bibnamefont
  {Golubitsky}}, \bibinfo {author} {\bibfnamefont {I.}~\bibnamefont {Stewart}},
  \ and\ \bibinfo {author} {\bibfnamefont {A.}~\bibnamefont {T{\"o}r{\"o}k}},\
  }\href@noop {} {\bibfield  {journal} {\bibinfo  {journal} {SIAM Journal on
  Applied Dynamical Systems}\ }\textbf {\bibinfo {volume} {4}},\ \bibinfo
  {pages} {78} (\bibinfo {year} {2005})}\BibitemShut {NoStop}%
\bibitem [{\citenamefont {Belykh}\ \emph {et~al.}(2000)\citenamefont {Belykh},
  \citenamefont {Belykh},\ and\ \citenamefont {Hasler}}]{belykh2000hierarchy}%
  \BibitemOpen
  \bibfield  {author} {\bibinfo {author} {\bibfnamefont {V.~N.}\ \bibnamefont
  {Belykh}}, \bibinfo {author} {\bibfnamefont {I.~V.}\ \bibnamefont {Belykh}},
  \ and\ \bibinfo {author} {\bibfnamefont {M.}~\bibnamefont {Hasler}},\
  }\href@noop {} {\bibfield  {journal} {\bibinfo  {journal} {Physical Review
  E}\ }\textbf {\bibinfo {volume} {62}},\ \bibinfo {pages} {6332} (\bibinfo
  {year} {2000})}\BibitemShut {NoStop}%
\bibitem [{\citenamefont {Belykh}\ \emph {et~al.}(2001)\citenamefont {Belykh},
  \citenamefont {Belykh},\ and\ \citenamefont {Mosekilde}}]{belykh2001cluster}%
  \BibitemOpen
  \bibfield  {author} {\bibinfo {author} {\bibfnamefont {V.~N.}\ \bibnamefont
  {Belykh}}, \bibinfo {author} {\bibfnamefont {I.~V.}\ \bibnamefont {Belykh}},
  \ and\ \bibinfo {author} {\bibfnamefont {E.}~\bibnamefont {Mosekilde}},\
  }\href@noop {} {\bibfield  {journal} {\bibinfo  {journal} {Physical Review
  E}\ }\textbf {\bibinfo {volume} {63}},\ \bibinfo {pages} {036216} (\bibinfo
  {year} {2001})}\BibitemShut {NoStop}%
\bibitem [{\citenamefont {Pecora}\ and\ \citenamefont
  {Carroll}(1998{\natexlab{b}})}]{Pe:Ca}%
  \BibitemOpen
  \bibfield  {author} {\bibinfo {author} {\bibfnamefont {L.}~\bibnamefont
  {Pecora}}\ and\ \bibinfo {author} {\bibfnamefont {T.}~\bibnamefont
  {Carroll}},\ }\href@noop {} {\bibfield  {journal} {\bibinfo  {journal} {Phys.
  Rev. Lett.}\ }\textbf {\bibinfo {volume} {80}},\ \bibinfo {pages} {2109}
  (\bibinfo {year} {1998}{\natexlab{b}})}\BibitemShut {NoStop}%
\bibitem [{\citenamefont {Kamei}\ and\ \citenamefont {Cock}(2013)}]{KameiCock}%
  \BibitemOpen
  \bibfield  {author} {\bibinfo {author} {\bibfnamefont {H.}~\bibnamefont
  {Kamei}}\ and\ \bibinfo {author} {\bibfnamefont {P.~J.~A.}\ \bibnamefont
  {Cock}},\ }\href@noop {} {\bibfield  {journal} {\bibinfo  {journal} {SIAM J.
  Applied Dynamical Systems}\ }\textbf {\bibinfo {volume} {12}},\ \bibinfo
  {pages} {352} (\bibinfo {year} {2013})}\BibitemShut {NoStop}%
\bibitem [{\citenamefont {MacArthur}\ and\ \citenamefont
  {Sanchez-Garcia}(2009)}]{MacArthur2009}%
  \BibitemOpen
  \bibfield  {author} {\bibinfo {author} {\bibfnamefont {B.}~\bibnamefont
  {MacArthur}}\ and\ \bibinfo {author} {\bibfnamefont {R.}~\bibnamefont
  {Sanchez-Garcia}},\ }\href@noop {} {\bibfield  {journal} {\bibinfo  {journal}
  {Physical Review E}\ }\textbf {\bibinfo {volume} {80}},\ \bibinfo {pages}
  {026117} (\bibinfo {year} {2009})}\BibitemShut {NoStop}%
\bibitem [{\citenamefont {MacArthur}\ \emph {et~al.}(2008)\citenamefont
  {MacArthur}, \citenamefont {Sanchez-Garcia},\ and\ \citenamefont
  {Anderson}}]{MacArthur2008}%
  \BibitemOpen
  \bibfield  {author} {\bibinfo {author} {\bibfnamefont {B.}~\bibnamefont
  {MacArthur}}, \bibinfo {author} {\bibfnamefont {R.}~\bibnamefont
  {Sanchez-Garcia}}, \ and\ \bibinfo {author} {\bibfnamefont {J.}~\bibnamefont
  {Anderson}},\ }\href@noop {} {\bibfield  {journal} {\bibinfo  {journal}
  {Discrete Appl. Math.}\ }\textbf {\bibinfo {volume} {156}},\ \bibinfo {pages}
  {3525} (\bibinfo {year} {2008})}\BibitemShut {NoStop}%
\bibitem [{\citenamefont {Group}(2005)}]{GAP4}%
  \BibitemOpen
  \bibfield  {author} {\bibinfo {author} {\bibfnamefont {T.~G.}\ \bibnamefont
  {Group}},\ }\href@noop {} {\emph {\bibinfo {title} {GAP: Groups, Algorithms,
  and Programming, Version 4.4}}}\ (\bibinfo  {publisher}
  {http://www.gap-system.org},\ \bibinfo {year} {2005})\BibitemShut {NoStop}%
\bibitem [{\citenamefont {Stein}(2013)}]{Stein}%
  \BibitemOpen
  \bibfield  {author} {\bibinfo {author} {\bibfnamefont {W.}~\bibnamefont
  {Stein}},\ }\href@noop {} {\emph {\bibinfo {title} {SAGE: Software for
  Algebra and Geometry Experimentation}}}\ (\bibinfo  {publisher}
  {http://www.sagemath.org/sage/, http://sage.scipy.org/},\ \bibinfo {year}
  {2013})\BibitemShut {NoStop}%
\bibitem [{\citenamefont {Poel}\ \emph {et~al.}(2015)\citenamefont {Poel},
  \citenamefont {Zakharova},\ and\ \citenamefont
  {Sch{\"o}ll}}]{poel2015partial}%
  \BibitemOpen
  \bibfield  {author} {\bibinfo {author} {\bibfnamefont {W.}~\bibnamefont
  {Poel}}, \bibinfo {author} {\bibfnamefont {A.}~\bibnamefont {Zakharova}}, \
  and\ \bibinfo {author} {\bibfnamefont {E.}~\bibnamefont {Sch{\"o}ll}},\
  }\href@noop {} {\bibfield  {journal} {\bibinfo  {journal} {Physical Review
  E}\ }\textbf {\bibinfo {volume} {91}},\ \bibinfo {pages} {022915} (\bibinfo
  {year} {2015})}\BibitemShut {NoStop}%
\bibitem [{\citenamefont {Manrubia}\ and\ \citenamefont
  {Mikhailov}(1999)}]{manrubia1999mutual}%
  \BibitemOpen
  \bibfield  {author} {\bibinfo {author} {\bibfnamefont {S.~C.}\ \bibnamefont
  {Manrubia}}\ and\ \bibinfo {author} {\bibfnamefont {A.~S.}\ \bibnamefont
  {Mikhailov}},\ }\href@noop {} {\bibfield  {journal} {\bibinfo  {journal}
  {Physical review E}\ }\textbf {\bibinfo {volume} {60}},\ \bibinfo {pages}
  {1579} (\bibinfo {year} {1999})}\BibitemShut {NoStop}%
\bibitem [{\citenamefont {Belykh}\ and\ \citenamefont
  {Hasler}(2011)}]{belykh2011mesoscale}%
  \BibitemOpen
  \bibfield  {author} {\bibinfo {author} {\bibfnamefont {I.}~\bibnamefont
  {Belykh}}\ and\ \bibinfo {author} {\bibfnamefont {M.}~\bibnamefont
  {Hasler}},\ }\href@noop {} {\bibfield  {journal} {\bibinfo  {journal} {Chaos:
  An Interdisciplinary Journal of Nonlinear Science}\ }\textbf {\bibinfo
  {volume} {21}},\ \bibinfo {pages} {016106} (\bibinfo {year}
  {2011})}\BibitemShut {NoStop}%
\bibitem [{\citenamefont {Hagerstrom}\ \emph {et~al.}(2012)\citenamefont
  {Hagerstrom}, \citenamefont {Murphy}, \citenamefont {Roy}, \citenamefont
  {H{\"o}vel}, \citenamefont {Omelchenko},\ and\ \citenamefont
  {Sch{\"o}ll}}]{Hagerstrom2012}%
  \BibitemOpen
  \bibfield  {author} {\bibinfo {author} {\bibfnamefont {A.~M.}\ \bibnamefont
  {Hagerstrom}}, \bibinfo {author} {\bibfnamefont {T.~E.}\ \bibnamefont
  {Murphy}}, \bibinfo {author} {\bibfnamefont {R.}~\bibnamefont {Roy}},
  \bibinfo {author} {\bibfnamefont {P.}~\bibnamefont {H{\"o}vel}}, \bibinfo
  {author} {\bibfnamefont {I.}~\bibnamefont {Omelchenko}}, \ and\ \bibinfo
  {author} {\bibfnamefont {E.}~\bibnamefont {Sch{\"o}ll}},\ }\href@noop {}
  {\bibfield  {journal} {\bibinfo  {journal} {Nature Physics}\ }\textbf
  {\bibinfo {volume} {8}},\ \bibinfo {pages} {658} (\bibinfo {year}
  {2012})}\BibitemShut {NoStop}%
\end{thebibliography}%

\end{document}